\documentclass[PRB,twocolumn,showpacs,preprintnumbers,prb]{revtex4}
\usepackage{amssymb}
\usepackage{graphics}
\usepackage{epsfig}

\begin{document}

\title{Colossal Positive Magnetoresistance in a Doped Nearly Magnetic
Semiconductor}
\author{Rongwei Hu$^{1,2}$, K. J. Thomas$^{1,\dagger }$, Y. Lee$^{3}$, T.
Vogt$^{1,\ast }$, E. S. Choi$^{4}$, V. F. Mitrovi\'{c}$^{2}$, R. P. Hermann$%
^{5,\S }$, F. Grandjean$^{5}$, P. C. Canfield$^{6}$, J. W. Kim$^{6}$ A. I.
Goldman$^{6}$ and C. Petrovic$^{1}$}
\affiliation{$^{1}$Condensed Matter Physics, Brookhaven National Laboratory, Upton, New
York 11973, USA\\
$^{2}$Physics Department, Brown University, Providence RI 02912, USA\\
$^{3}$Department of Earth System Sciences, Yonsei University, Seoul 120749,
Korea\\
$^{4}$NHMFL/Physics, Florida State University, Tallahassee, Florida 32310,
USA\\
$^{5}$Department of Physics, Universit\'{e} de Li\`{e}ge, Belgium\\
$^{6}$Ames Laboratory and Department of Physics and Astronomy, Iowa State
University, Ames, IA 50011, USA }
\date{\today }

\begin{abstract}
We report on a positive colossal magnetoresistance (MR) induced by
metallization of FeSb$_{2}$, a nearly magnetic or "Kondo" semiconductor with 
\textit{3d} ions. We discuss contribution of orbital MR and quantum
interference to enhanced magnetic field response of electrical resistivity.
\end{abstract}
\maketitle

There is at present considerable technological interest in the
magnetoresistive effect that is central to the operation of devices in
magnetic storage media and a wide variety of magnetic sensors. The desire to
maximize this effect has raised interest in new materials and mechanisms
associated with the large change in electrical resistance in magnetic field.%
\cite{Anisimov1} FeSb$_{2}$ is a narrow gap semiconductor whose magnetic
properties strongly resemble nearly magnetic or "Kondo" insulator FeSi.\cite%
{ja}$^{,}$\cite{ja2} The \textit{ab initio} LDA+\textit{U} electronic
calculations found that the ground state of FeSb$_{2}$ is nearly
ferromagnetic.\cite{Moris} Temperature induced paramagnetic moment for field
applied parallel to $\widehat{b}$ - axis coincides with increased conduction
at high temperatures.\cite{ja}$^{,}$\cite{ja2}$^{,}$\cite{Rongwei}
Electrical transport measurements showed pronounced anisotropy, metallic
conductivity above 40 K and activated below that temperature for current
applied along $\widehat{b}$ axis. The exact temperature of metal -
semiconductor crossover was found to be very sensitive to small current
misalignment, implying quasi - one dimensional nature of electronic
transport.\cite{ja} Optical conductivity revealed anisotropic energy gap E$%
_{g}$ in the spectral range between 100 - 350 cm$^{-1}$ and negligible Drude
weight of $\sigma (\omega )$ at low frequencies, i.e. a true insulating
state. In addition, a full recovery of spectral weight occurs only above 1
eV, suggesting contribution of larger energy scales than in conventional MIT
transitions where thermal excitations of charge carriers through E$_{g}$
brings about redistribution of the spectral weight just above the gap.\cite%
{Leo} Metal-insulator transitions in FeSi have been induced by band narrowing%
\cite{Sunmog} or by insertion of carriers,\cite{Dave} albeit with modest
magnetoresistance. Unlike Fe$_{1-x}$Co$_{x}$Si, CMR is observed for all $%
x\leq $ 0.4 in Fe$_{1-x}$Co$_{x}$Sb$_{2}$.

Fe$_{1-x}$Co$_{x}$Sb$_{2}$ single crystals were grown from excess Sb flux.%
\cite{ja} X-ray powder diffraction experiments and crystal orientation were
performed at the beamlines X7A and X22C of the NSLS at the Brookhaven
National Laboratory. Electrical transport and transverse magnetoresistance
(MR) measurements were carried out in Quantum Design MPMS and PPMS
instruments. For Hall resistivity $\rho _{xy}$, the current was applied
along the highly conducting $\widehat{b}$ - axis and voltage was picked up
in the orthogonal (Hall) [101] direction. To completely cancel out the
longitudinal voltage contribution, magnetic field was swept from H = 9 T to
H = -9 T and one half of the voltage difference was taken as the Hall
voltage. The high field transverse MR measurements were carried out at the
NHMFL in Tallahassee, FL up to 300 kOe. The powder X-ray patterns show that
the Co substitution on the Fe site uniformly contracts the unit cell volume,
in agreement with Vegard's law.\cite{Rongwei} The M\"{o}ssbauer data
indicate that there is no localized magnetism on the Fe site and no visible
subspectrum of impurities.

\begin{figure}[tbp]
%%%%%%%%%%%%%%%%%%%   F I G U R E  1  %%%%%%%%%%%%%%%%%
\centerline{\includegraphics[height=4.5in]{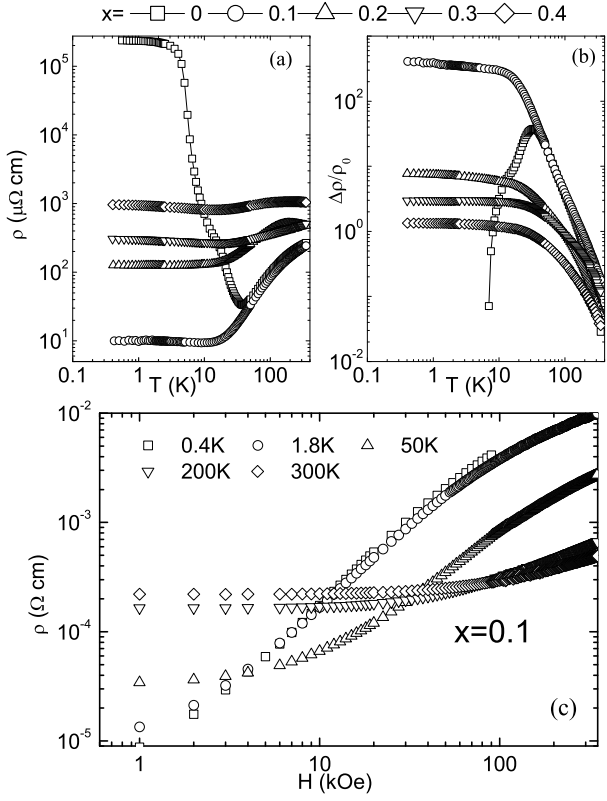}} %%%%%%%%%
%%%%%%%%%%%%%%%%%%%%%%%%%%%%%%%%%%%%%%%%%%%%%
\vspace*{-0.2cm}
\caption{(a) Electrical transport properties of {Fe}$_{{1-x}}${Co}$_{{x}}${Sb%
$_{2}$ } dramatically change with small change in shoichiometry. {The sample
with }${x=0.1}${\ manifests lowest resistivity. With further Co substitution
we see gradual increase in residual resistivity. (b) Temperature dependence
of magnetoresistance }$MR$=[$\protect\rho $(90kOe)-$\protect\rho $(0)]/$%
\protect\rho $(0){\ in doped Fe$_{1-x}$Co$_{x}$Sb$_{2}$ semiconductor alloys
for }${x=0-0.4}${. CMR is observed for all }${x}${. (c) Resistivity
isotherms of Fe$_{0.9}${Co}$_{0.1}${Sb}$_{2}$ up to 350 kOe. Resistivity
increases three orders of magnitude (103,100\%) at 1.8K.}}
\end{figure}

\begin{figure}[t]
%%%%%%%%%%%%%%%%%%%   F I G U R E  2  %%%%%%%%%%%%%%%%%
\centerline{\includegraphics[height=4.5in]{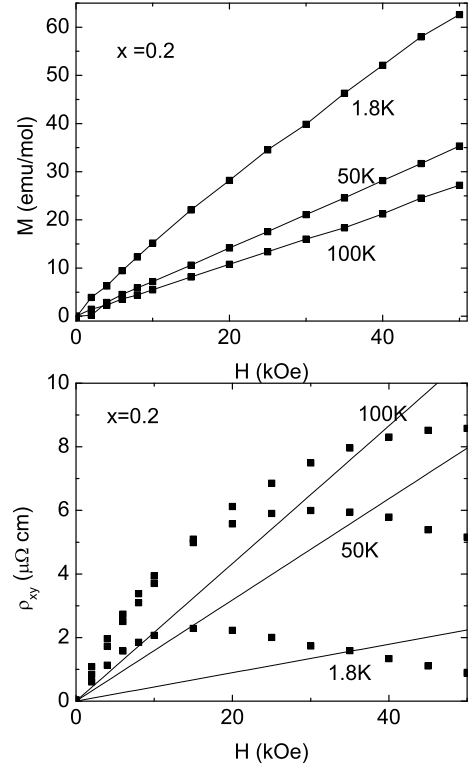}} %%%%%%%%%
%%%%%%%%%%%%%%%%%%%%%%%%%%%%%%%%%%%%%%%%%%%%%
\vspace*{-0.2cm}
\caption{Hall resistivity versus magnetic field. Solid lines are the fits
when magnetization M is taken into account.}
\end{figure}

The application of a magnetic field of 90 kOe induces up to 2.5 orders of
magnitude change in the resistivity in Fe$_{1-x}$Co$_{x}$Sb$_{2}$ (x = 0 -
0.4) and 30\% of MR at room temperature for $x$ = 0.1 (Fig. 1). This is
comparable to MR in the colossal magnetoresistive manganites.\cite{Salamon}
What could be the most likely cause of CMR in this system? The mechanism
could involve a presence of charge carriers from different parts of the
Fermi surface that have different scattering times or a breakdown of the
semiclassical transport theory and relation between conductivity and
scattering time $\rho \sim 1/\tau $.\cite{SrRuO3}

Hall resistivity is shown as a nonlinear function of magnetic field in Fig.
2(b). The pronounced field dependence of Hall constant is reminiscent of
anomalous Hall effect. When there are magnetic moments involved, the Hall
resistivity can be written as \cite{Handley} 
\[
\rho _{xy}(H)=R_{0}H+R_{s}M(H) 
\]

where R$_{0}$ and R$_{s}$ are the normal and spontaneous Hall constants, and
M is the sample magnetization. Fitting $\rho _{xy}$ data by using the
experimental values of M(H) (Fig. 2(a)), we found that the anomalous Hall
effect is unlikely cause of Hall resistivity nonlinearity. The nonlinearity
of Hall resistivity in field and band structure calculations\cite{Moris}
suggest that there are more than one band participating in the conduction
(Fig. 3).

In what follows we discuss the case when there are more than one type of
carrier participating in the conduction. The Hall constant takes the general
form\cite{Finkman}%
\begin{eqnarray*}
R_{H} &=&-\frac{1}{H}\frac{\sum \sigma _{xy}^{i}}{(\sum \sigma
_{xx}^{i})^{2}+(\sum \sigma _{xy}^{i})^{2}} \\
\sigma _{xx}^{i} &=&\frac{qn_{i}\mu _{i}}{1+\mu _{i}^{2}H^{2}},\text{ }%
\sigma _{xy}^{i}=\frac{qn_{i}\mu _{i}^{2}H}{1+\mu _{i}^{2}H^{2}}
\end{eqnarray*}

where $\sigma _{xx}^{i}$ and $\sigma _{xy}^{i}$are longitudinal and Hall
conductivities of individual bands, $\mu _{i}$ is the band mobility. A
matrix formalism for the Hall effect in multicarrier semiconductor systems
was devised and provided a closed form formula for two or three-carrier
systems.\cite{JSKim1}$^{,}$\cite{JSKim2} The magnetoresistance within the
same formalism for two-carrier system is:

\[
R_{H}=\rho _{0}\frac{\alpha _{2}+\beta _{2}H^{2}}{1+\beta _{3}H^{2}},MR_{D}=%
\frac{\alpha _{D}H^{2}}{1+\beta _{D}H^{2}} 
\]

$\alpha _{2}=f_{1}\mu _{1}+f_{2}\mu _{2},\beta _{2}=(f_{1}\mu _{2}+f_{2}\mu
_{1})\mu _{1}\mu _{2}$

$\beta _{3}=\beta _{D}=(f_{1}\mu _{2}+f_{2}\mu _{1})^{2},$ $\alpha
_{D}=f_{1}f_{2}(\mu _{1}-\mu _{2})^{2}$

and for three-carrier systems:

\begin{center}
\begin{eqnarray*}
\text{\ \ \ \ }R_{H} &=&\rho _{0}\frac{\alpha _{2}+\beta _{2}H^{2}+\gamma
_{2}H^{4}}{1+\beta _{3}H^{2}+\gamma _{3}H^{4}} \\
MR_{T} &=&\frac{(\alpha _{T}+\gamma _{T}H^{2})H^{2}}{1+(\beta _{T}+\delta
_{T}H^{2})H^{2}}
\end{eqnarray*}
\end{center}

$\alpha _{2}=f_{1}\mu _{1}+f_{2}\mu _{2}+f_{3}\mu _{3}$

$\beta _{2}=f_{1}\mu _{1}(\mu _{2}^{2}+\mu _{3}^{2})+f_{2}\mu _{2}(\mu
_{1}^{2}+\mu _{3}^{2})+f_{3}\mu _{3}(\mu _{1}^{2}+\mu _{2}^{2})$

$\beta _{3}=\beta _{T}=(f_{1}\mu _{2}+f_{2}\mu _{1})^{2}+(f_{2}\mu
_{3}+f_{3}\mu _{2})^{2}+(f_{1}\mu _{3}+f_{3}\mu _{1})^{2}+2(f_{1}f_{2}\mu
_{3}^{2}+f_{2}f_{3}\mu _{1}^{2}+f_{1}f_{3}\mu _{2}^{2})$

$\gamma _{2}=(f_{1}\mu _{2}\mu _{3}+f_{2}\mu _{1}\mu _{3}+f_{3}\mu _{1}\mu
_{2})\mu _{1}\mu _{2}\mu _{3}$

$\gamma _{3}=\delta _{T}=(f_{1}\mu _{2}\mu _{3}+f_{2}\mu _{1}\mu
_{3}+f_{3}\mu _{1}\mu _{2})^{2}$

$\alpha _{T}=f_{1}f_{2}(\mu _{1}-\mu _{2})^{2}+f_{1}f_{3}(\mu _{1}-\mu
_{3})^{2}+f_{2}f_{3}(\mu _{2}-\mu _{3})^{2}$

$\beta _{T}=(f_{1}\mu _{2}+f_{2}\mu _{1})^{2}+(f_{3}\mu _{1}+f_{1}\mu
_{3})^{2}+(f_{2}\mu _{3}+f_{3}\mu _{2})^{2}$

$\gamma _{T}=f_{1}f_{2}(\mu _{1}-\mu _{2})^{2}\mu _{3}^{2}+f_{1}f_{3}(\mu
_{1}-\mu _{3})^{2}\mu _{2}^{2}+f_{2}f_{3}(\mu _{2}-\mu _{3})^{2}\mu _{1}^{2}$

\begin{figure}[t]
%%%%%%%%%%%%%%%%%%%   F I G U R E  3  %%%%%%%%%%%%%%%%%
\centerline{\includegraphics[height=2.8in]{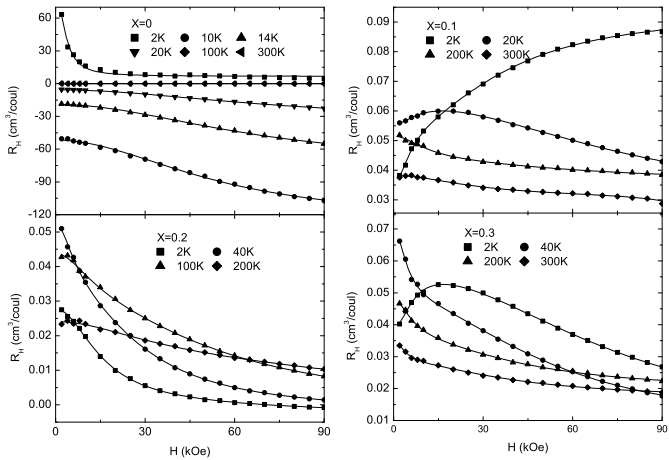}} %%%%%%%%%
%%%%%%%%%%%%%%%%%%%%%%%%%%%%%%%%%%%%%%%%%%%%%
\vspace*{-0.2cm}
\caption{(a) Hall constant R$_{H}$ = $\protect\rho _{xy}$/H of FeSb$_{2}$ is
well described by a two-carrier model. (b)-(d) For Fe$_{1-x}$Co$_{x}$Sb$_{2}$
(0.1$\leq x\leq $0.4) the fits of two-carrier model to experimental data are
rather unsatisfactory since it cannot account for maxima and minima in R$%
_{H} $(H). Co substitution in Fe$_{1-x}$Co$_{x}$Sb$_{2}$ fills energy bands
that are not involved in thermal excitation of carriers across the gap in
FeSb$_{2}$. R$_{H}$ shows excellent agreement with the three-carrier model
for all $x$, selected data are shown for clarity. Solids lines are fits to
the multicarrier model.}
\end{figure}

\begin{figure}[t]
%%%%%%%%%%%%%%%%%%%   F I G U R E  4  %%%%%%%%%%%%%%%%%
\centerline{\includegraphics[height=2.8in]{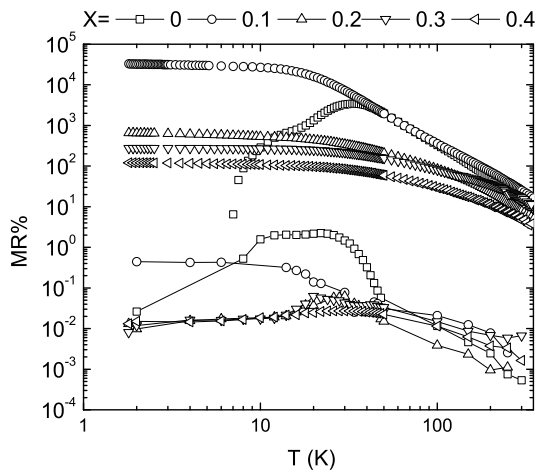}} %%%%%%%%%
%%%%%%%%%%%%%%%%%%%%%%%%%%%%%%%%%%%%%%%%%%%%%
\vspace*{-0.2cm}
\caption{Magnetoresistance calculated according to multicarrier model for Fe$%
_{1-x}$Co$_{x}$Sb$_{2}$ in 90 kOe shown in the low plot as linked symbols,
compared to the observed MR shown as the scattered symbols.}
\end{figure}

where $\rho _{0}$ is the zero field resistivity, $\mu _{i}$ is the mobility
of i$^{th}$ carrier and $f_{i}=|n_{i}\mu _{i}|/\sum |n_{i}\mu _{i}|$ is the
f factor, $n_{i}$ is the carrier concentration, as defined in Ref. 13. The
agreement with our experimental values of R$_{H}$ was excellent as the solid
lines shown in Fig. 3. FeSb$_{2}$ can be described well by two-carrier
model. Fe$_{1-x}$Co$_{x}$Sb$_{2}$ is best described within three-carrier
model since two carrier model cannot explain minima and maxima in R$_{H}$ = $%
\rho _{xy}$/H, therefore only three carrier fits are shown in Fig. 3(b)-(d).
Using the obtained fitting parameters for the Hall constant, we can now
solve numerically for mobilities of individual carriers and MR. We show the
result in Fig. 4: the calculated orbital MR is approximately four orders of
magnitude smaller than the observed MR for all $x$. This suggests that the
conventional orbital mechanism can not account for the large
magnetoresistance in Fe$_{1-x}$Co$_{x}$Sb$_{2}$. Similar shape of calculated
and measured MR, taken together with Fig. 3, argues in favor of the validity
of the multiband electronic transport. In what follows we show that the
experimental MR is amplified by quantum interference effects.

Disordered metallic Fe$_{1-x}$Co$_{x}$Sb$_{2}$ alloys are derived from an
insulator with strong correlation effects.\cite{Moris} Therefore, possible
mechanism for enhanced positive MR could involve change of electronic
structure induced by Co and quantum interference effects through
contribution of electron - electron interaction enhanced by disorder.\cite%
{Lee}$^{,}$\cite{Dugdale} Fig. 5(a) shows qualitatively that contribution of
Coulomb interaction in the presence of random impurity scattering may be
important. More information is obtained from magnetic field dependence of
positive MR isotherms taken at 1.8 K, shown in Fig. 5(b). In the limit of
strong spin-orbit scattering, correction to quadratic MR along the high
conductivity axis in Fe$_{1-x}$Co$_{x}$Sb$_{2}$ is expressed by:

\begin{table*}[t]
\caption{Parameters of the fits to quantum correction of magnetoresistance. $%
\protect\alpha F$ is the proportional parameter in the Coulomb interaction
contribution. L$_{f}$ is the phase coherence length, b is the 1D conduction
channel width and c is the coefficient of classical quadratic term.}%
\begin{tabular}{p{1in}p{1in}p{1in}p{1in}p{1in}}
\hline\hline
$x$ & $\alpha F$ & $L_{f}(nm)$ & $b(nm)$ & $c$ \\ \hline
$0$ & $1.2\times 10^{-9}$ & $162$ & $1.6$ & $0.001$ \\ 
$0.1$ & $6.2\times 10^{-7}$ & $1463$ & $0.2$ & $0.055$ \\ 
$0.2$ & $5.1\times 10^{-9}$ & $534$ & $2.8$ & $0.004$ \\ 
$0.3$ & $7.3\times 10^{-9}$ & $188$ & $4.0$ & $0.002$ \\ 
$0.4$ & $9.9\times 10^{-9}$ & $128$ & $5.1$ & $0.001$ \\ \hline\hline
\end{tabular}%
\end{table*}

\begin{figure}[tbp]
%%%%%%%%%%%%%%%%%%%   F I G U R E  5  %%%%%%%%%%%%%%%%%
\centerline{\includegraphics[height=4.8in]{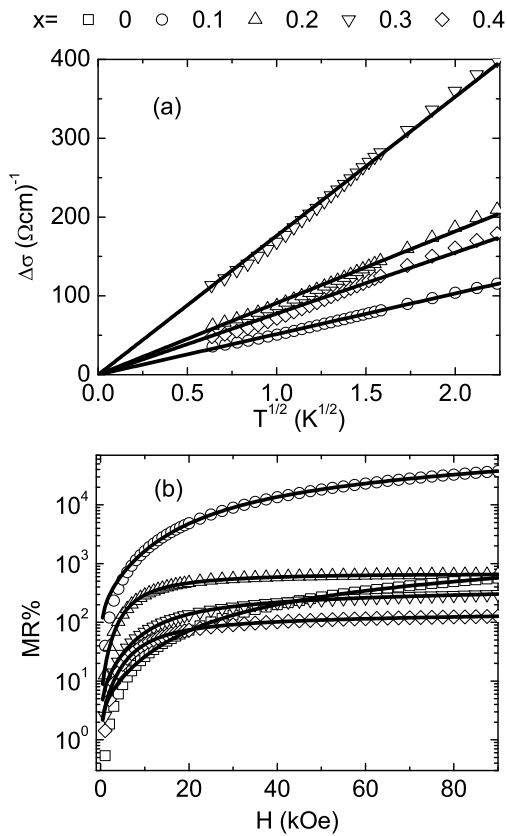}} %%%%%%%%%
%%%%%%%%%%%%%%%%%%%%%%%%%%%%%%%%%%%%%%%%%%%%%
\vspace*{-0.2cm}
\caption{(a) Magnetoconductivity of Fe$_{1-x}$Co$_{x}$Sb$_{2}$. Conductivity
change $\Delta \protect\sigma =\protect\sigma (H,T)-\protect\sigma (H,0)$
with $\protect\sigma (H,0)$ determined from fits of data to a T$^{1/2}$
dependence at $H=90$ $kOe.$ (b) Quasi 1D weak localization fits to MR at T =
1.8 K for Fe$_{1-x}$Co$_{x}$Sb$_{2}$. Data for $x$ = 0 are shown at T = 100
K in metallic quasi - 1D regime of $\widehat{b}$ - axis conductivity. }
\end{figure}

\begin{eqnarray*}
MR &=&1/\rho _{0}0.77\alpha F(g\mu _{B}/k_{B})^{1/2}H^{1/2} \\
&&-\frac{\rho _{0}e^{2}L_{f}}{\pi \hbar b^{2}}[(1+\frac{b^{2}L_{f}^{2}}{%
12L_{B}^{4}})^{-1/2}-1]+cH^{2}
\end{eqnarray*}%
where $L_{f}$ is the phase coherence length, $L_{B}=\sqrt{\hbar /2eB}$ is
the magnetic length, $b$ is the width of quasi 1D channel, and $F$ is from
Hartree interaction. The first term is the contribution of Coulomb
interaction\cite{Morita}$^{,}$\cite{PALee} and the second term is the
contribution of weak localization of quasi 1D model.\cite{AltshulerBL}$^{,}$%
\cite{Yakimov} Magnetoresistance is positive, as expected for the case of
strong spin-orbit scattering in nearly magnetic conductors such as Pd and Pt
alloys, but also FeSb$_{2}$.\cite{Moris}$,$\cite{Millis} The fitting
parameters are listed in Table I. As we can see, the contribution of Coulomb
interaction to MR is of the order of several percent and the weak
localization is dominant. Two inequalities justify our quasi 1D model: 1)
the fitting parameters satisfy the criterion for the interference correction
to be of \ a 1D character: $b<L_{f}$ and 2) in a field of 9T, the magnetic
length $L_{B}$ is about 6nm, therefore $b<L_{B}$ is satisfied, otherwise the
system should behave as three dimensional. The calculated MR agrees well
with the observed MR. At T = 1.8 K, the calculated MR in an H = 90 kOe \
field is about $34,300\%$ for Fe$_{0.9}$Co$_{0.1}$Sb$_{2}$, as compared to
the observed value of $36,088\%$. The value is derived using a measured
carrier concentration $n=8.5\times 10^{19}$ $cm^{-3}$ and a residual
resistivity $\rho _{0}=1.0\times 10^{-5}$ $\Omega cm$ of Fe$_{0.9}$Co$_{0.1}$%
Sb$_{2}$, and $L_{f}\simeq 240L_{B}$. The width of the quasi 1D channel $b$
is of the order of the unit cell. Quasi 1D localization effects are observed
in pure FeSb$_{2}$ at 100K (Fig. 5(b)). It implies that the electronic
transport is dominated by the singular corrections at the density of states
at unusually high temperatures, apparently without usual cutoff by thermal
effects, probably due to high spin orbit scattering rate 1/$\tau $s .\cite%
{Millis} Similar situation has been observed in Ta$_{4}$Te$_{4}$Si at 15K%
\cite{Stolovits} and Si nanowires at 27K.\cite{Dayen}

In conclusion, we have shown CMR effect in doped nearly magnetic
semiconductor FeSb$_{2}$. Application of magnetic fields up to 350 kOe
results in 103,100\% increase in the resistivity at 1.8 K and 124\% increase
at room temperature. Our results suggest contribution of multiple electronic
bands to electrical conduction. Quantum interference arising from Coulomb
interactions and weak localization in the presence of strong spin-orbit
scattering is the dominant mechanism of CMR.

We thank Maxim Dzero, Sergey Bud'ko, Zachary Fisk and T. Maurice Rice for
useful communication.

This work was carried out at the Brookhaven National Laboratory which is
operated for the U.S. Department of Energy by Brookhaven Science Associates
(DE-Ac02-98CH10886). A portion of this work was performed at the National
High Magnetic Field Laboratory, which is supported by NSF Cooperative
Agreement No. DMR-0084173, by the State of Florida, and by the U.S.
Department of Energy. This work was also supported in part by the National
Science Foundation DMR-0547938 (V. F. M.).

$^{\dagger}$Present address: Nature Publishing Group, London, UK

$^{\ast }$Present address: Department of Chemistry and Biochemistry,
University of South Carolina, Columbia, SC 29208

$^{\S }$Present address: Institut f\"{u}r Festk\"{o}rperforschung,
Forschungzentrum J\"{u}lich GmbH, D-52425 J\"{u}lich, Germany


\begin{thebibliography}{99}
\bibitem{Anisimov1} V. I. Anisimov, R. Hlubina, M. A. Korotin, V. V.
Mazurenko, T. M. Rice, A. O. Shorikov, and M. Sigrist, Phys. Rev. Lett. 
\textbf{89}, 257203 (2002)

\bibitem{ja} C. Petrovic, J. W. Kim, S. L. Bud'ko, A. I. Goldman, and P. C.
Canfield, Phys. Rev. B \textbf{67}, 155205 (2003)

\bibitem{ja2} C. Petrovic, Y. Lee, T. Vogt, N. Dj. Lazarov, S. L. Bud'ko,
and P. C. Canfield, Phys. Rev. B \textbf{72}, 045103 (2005)

\bibitem{Moris} A. V. Lukoyanov, V.V. Mazurenko, V.I. Anisimov, M. Sigrist
and T.M. Rice, European Physical Journal B \textbf{53}, 205 (2006)

\bibitem{Rongwei} Rongwei Hu, V. F. Mitrovic and C. Petrovic, Phys. Rev. B 
\textbf{74}, 195130 (2006)

\bibitem{Leo} A. Perucchi, L. Degiorgi, Rongwei Hu, C. Petrovic and V.F.
Mitrovic, European Physical Journal B \textbf{54}, 175 (2006)

\bibitem{Sunmog} S. Yeo, S. Nakatsuji, A. D. Bianchi, P. Schlottmann, Z.
Fisk, L. Balicas, P. A. Stampe, and R. J. Kennedy, Phys. Rev. Lett. \textbf{%
91}, 046401 (2003)

\bibitem{Dave} N. Manyala, Y. Sidis, J. F. DiTusa, G. Aeppli, D.P. Young and
Z. Fisk, Nature \textbf{404}, 581 (2000)

\bibitem{Salamon} Myron B. Salamon and Marcelo Jaime, Rev. Mod. Physics 
\textbf{73}, 583 (2001)

\bibitem{SrRuO3} J. S. Dodge, C. P. Weber, J. Corson, J. Orenstein, Z.
Schlesinger, J. W. Reiner, and M. R. Beasley, Phys. Rev. Lett. \textbf{85},
4932 (2000)

\bibitem{Handley} R.C. O'Handley, in The Hall Effect and its Application,
edited by C.L. Chien and C.R. Westgate (Plenum, New York,1980), p. 417.

\bibitem{Finkman} E. Finkman and Y. Nemirovsky, J. Appl. Phys. 53, 1055
(1982)

\bibitem{JSKim1} J.S. Kim, J. App. Phys. \textbf{86}, 3187 (1999)

\bibitem{JSKim2} J.S. Kim, J. App. Phys. \textbf{84}, 292(1998)

\bibitem{Lee} Patrick A. Lee and T. \ V. Ramakrishnan, Rev. Mod. Phys. 
\textbf{57}, 287 (2005)

\bibitem{Dugdale} J. S. Dugdale, \textit{The Electrical Properties of
Disordered Metals} (Cambridge University Press, New York 1995)

\bibitem{Morita} S. Morita,\textit{\ } Y. Isawa, T. Fukase, S. Ishida, Y.
Koike, Y. Takeuti and N. Mikoshiba, Phys. Rev. B \textbf{25}, 5570 (1982)

\bibitem{PALee} P. A. Lee and T. V. Ramakrishnan, Phys. Rev. B \textbf{26},
4009 (1982)

\bibitem{AltshulerBL} B. L. Al'tshuler and A. G. Aronov, JETP Lett. \textbf{%
33}, 499 (1981)

\bibitem{Yakimov} A. I. Yakimov and A. V. Dvurechenskii, JETP Lett. \textbf{%
69}, 202 (1999)

\bibitem{Millis} A. J. Millis and P. A. Lee, Phys. Rev. B 30, 6170 (1984)

\bibitem{Stolovits} A. Stolovits, A. Sherman, K. Ahn and R. K. Kremer, Phys.
Rev. B, \textbf{62} 10565 (2000)

\bibitem{Dayen} J.F. Dayen, A. Rumyantseva, C. Ciornei, T. L. Wade, and J.E.
Wegrowe, Pribat and C. Sorin Cojocaru, App. Phys. Lett. \textbf{90}, 173110
(2007)
\end{thebibliography}
\end{document}